\documentclass[preprint,12pt]{elsarticle}
\biboptions{sort&compress}
\usepackage[utf8]{inputenc}
\usepackage{graphicx}
\usepackage{amsmath}
\usepackage{xfrac}
\DeclareMathOperator*{\argmin}{argmin}

\usepackage[table,xcdraw]{xcolor} 

\usepackage{gensymb}

\usepackage[nolist]{acronym}

\usepackage{braket}

\usepackage{physics}

\usepackage{amssymb,mathtools}
\usepackage{subcaption}

\usepackage[font=small,labelfont=bf]{caption}
\usepackage[font=small,labelfont=bf]{subcaption}

\usepackage{hyperref}

\usepackage{amsfonts}
\usepackage{nicefrac}

\usepackage{algorithm} 
\usepackage{algpseudocode}
\usepackage{cite}

\usepackage[final]{pdfpages}

\usepackage{pgfplots}
\pgfplotsset{compat=newest}
\pgfplotsset{plot coordinates/math parser=false} 

\usepackage{tabu}

\newtheorem{definition}{Definition}

\newtheorem{lemma}{Lemma}

\newtheorem{initialization}{Initialization}

\providecommand{\openbox}{\leavevmode
  \hbox to.77778em{%
  \hfil\vrule
  \vbox to.675em{\hrule width.6em\vfil\hrule}%
  \vrule\hfil}}
\makeatletter
\DeclareRobustCommand{\qed}{%
  \ifmmode
    \eqno \def\@badmath{$$}
    \let\eqno\relax \let\leqno\relax \let\veqno\relax
    \hbox{\openbox}%
  \else
    \leavevmode\unskip\penalty9999 \hbox{}\nobreak\hfill
    \quad\hbox{\openbox}%
  \fi
}
\makeatother
\newtheorem{thm}{Theorem}

\newdefinition{rmk}{Remark}
\newproof{pf}{Proof}
\newproof{pot}{Proof of Theorem \ref{thm2}}
\newcommand{\mt}{m_{\mathcal{T}}}
\newcommand{\Et}{\mathbb{E}_\mathcal{T}}
\newcommand{\T}{\mathcal{T}}
\newcommand{\X}{\mathcal{X}}
\newcommand{\IntSet}{\mathbb{Z}}
\newcommand{\R}{\mathbb{R}}
\newcommand{\A}{\mathcal{A}}
\newcommand{\del}{\partial}
\newcommand{\C}{\mathcal{C}}

\newcommand{\Ls}{\mathcal{L}}
\newcommand{\I}{\mathbb{I}}

\usepackage{mathtools}

\usepackage{textcomp}

\journal{Systems \& Control Letters}

\begin{document}
\begin{frontmatter}

\title{Iterative, Small-Signal \texorpdfstring{$\mathcal{L}_2$}{L2} Stability Analysis of Nonlinear  Constrained Systems}

\author[Duke]{Reza Lavaei}
\ead{reza.lavaei@duke.edu}
\author[Duke]{Leila J. Bridgeman\corref{cor1}}
\ead{leila.bridgeman@duke.edu}

\cortext[cor1]{Corresponding author}
\affiliation[Duke]{organization={Thomas Lord Depertment of Mechanical Engineering and Materials Science, Duke University},
            addressline={304 Research Drive}, 
            city={Durham},
            postcode={27708}, 
            state={NC},
            country={USA}}

\begin{abstract}
This paper provides a method to analyze the small-signal $\Ls_2$-gain of control-affine nonlinear systems on compact sets via iterative semi-definite programs. First, a continuous piecewise affine storage function and the corresponding upper bound on the $\Ls_2$-gain are found on a bounded, compact set's triangulation. Then, to ensure that the state does not escape this set, a continuous piecewise affine barrier function is found that is robust to small-signal inputs. Small-signal $\Ls_2$ stability then holds inside each sublevel set of the barrier function inside the set where the storage function was found. The method's effectiveness is shown in a numerical example.
\end{abstract}

\begin{highlights}
\item Theorem 3 establishes criteria to bound the $\mathcal{L}_2$ gain of a nonlinear system on a robustly positive invariant subset of the state-space. It simultaneously identifies a bound on the input disturbance signal's infinity norm to which the subset is robustly positive invariant.
\item Algorithms are proposed to evaluate the criteria of Theorem 3, relying only on repeated solution of well-posed, convex optimization problems.
\item The paper provides systematic means to evaluate the gain of constrained, nonlinear systems under a small-signal requirement.
\item An iterative method allows rigorous verification of the small-signal requirement.
\end{highlights}

\end{frontmatter}

\section{Introduction}
\label{sec:introduction}
Considering a system as an input-output map between two normed vector spaces, the $\Ls_2$-gain bounds the norm of the input above in the energy sense as it passes through the system \citep{boyd1987comparison}. If it is finite, input-output stability is established in the sense of $\Ls_2$. This analysis is important in evaluating the performance of stable systems in the presence of norm-bounded disturbances \citep{zames1981optimalsensitivity,zames1983minmax}, and in studying the stability of interconnected systems via the Small-Gain Theorem \citep{zames1966input}, which underpins $\mathcal{H}_\infty$ robust control. This paper gives an iterative method to bound the $\Ls_2$-gain of nonlinear time-invariant maps with state constraints. 

For stable, linear, time-invariant systems, the $\Ls_2$-gain is equivalent to the $\mathcal{H}_\infty$ norm of the corresponding transfer function due to Parseval's Theorem \citep[p. 51]{zhou1998essentials} and can be found using the Bounded Real Lemma \citep{doyle1988state}. The connection between the frequency and time domains allowed an efficient bisection algorithm in \citep{boyd1989bisection,bruinsma1990fast} improving the method of \citep{guo1988recursive} that was purely based on frequency domain analysis. These works inspired future, more sophisticated methods of \citep{scherer1990h,lin1999computation,gahinet1992numerical,gattami2015simple,benner2018,xia2020sector}.

For nonlinear systems, $\Ls_2$-gain analysis remains a challenge \citep{van19922,ball1996viscosity,helton1999extending}. An upper-bound on $\Ls_2$-gain is typically sought and tight bounds are hard to establish \citep{zhang2012performance}. Due to modeling or physical limitations, and safety considerations, the input-output mapping may only be defined for a subset of the state-space. This further complicates $\Ls_2$-gain analysis, even for linear systems. That said, the \ac{HJ} and dissipativity inequalities can be used to analyze $\Ls_2$-gain for nonlinear systems; it is just challenging to establish the existence of a positive semi-definite storage function satisfying these inequalities. Additionally, one needs to make sure that for a subset of admissible inputs, the \ac{HJ} inequality holds and that the state remains inside the set of admissible states. In \citep{summers2013quantitative}, these requirements were verified for polynomial systems with polynomial storage functions. However, there is no assurance that the storage functions must be polynomial, and for general nonlinear systems, finding a storage function that satisfies the \ac{HJ} inequality is generally a non-convex, computationally challenging problem. 

Instead of seeking out polynomial storage functions, this paper analyses small-signal $\Ls_2$-gain of constrained, time-invariant, nonlinear maps by seeking \ac{CPA} storage functions on triangulations of the set of admissible states. This allows the \ac{HJ} inequalities to be verified everywhere on a set while only completing calculations at finitely many points. Triangulation refinement searches a rich class of possible storage functions, and where polynomial dynamics were needed to formulate a tractable search problem with polynomial storage functions in \citep{summers2013quantitative}, the search for \ac{CPA} storage functions can be accomplished through the iterative solution of \acp{SDP} for more general nonlinear dynamics. The techniques of \citep{giesl2012,gieslRevCPA2013,giesl2015}, which use \ac{CPA} Lyapunov functions to establish Lyapunov stability, provide inspiration for the proposed search. However, several adjustments are needed to seek out storage functions instead of Lyapunov functions, and to bound $\Ls_2$-gain, rather than simply establish stability. Further, it is crucial here to determine for what inputs the system remains in the finite region where analysis was performed. This necessitates an additional search for barrier-like functions to establish set invariance. Thus, to introduce a method for bounding the $\Ls_2$-gain of constrained, input-affine maps, several contributions have been made:
    \begin{itemize}
        \item Providing sufficient conditions to ensure satisfaction of the HJ inequality over a compact set through modified calculations on a finite set of points. 
        \item Providing sufficient conditions to verify the existence of compact robust invariant sets through calculations on a finite set of points. This entails modified barrier function requirements that relax standard decrease conditions inside a set. 
        \item Exploiting the above to use CPA functions as a rich, flexible search class of storage and barrier functions.
        \item For nonlinear, input-affine dynamics, formulating the search for a storage function, barrier function, $\Ls_2$-gain bound, and the set of inputs for which small-signal properties hold as \ac{SDP} sequences that can be solved efficiently through convex optimization.
    \end{itemize}

\section{Preliminaries}
\textbf{Notation.} The induced infinity and two norms of a matrix, $A$, will respectively be denoted $||A||_\infty = \sup_{||v||_\infty = 1}||Av||_{\infty}$ and $||A||_2 = \sup_{||v||_2 = 1}||Av||_2$. The $p$-norm, where $p\in[0,\infty]$, of $u\in\R^m$ is denoted by $||u||_p$. The normed space of 
functions $u:[0,\infty]\mapsto\R^m$ is denoted by $\Ls_p$ with the norm $||u||_{\Ls_p}=(\int_0^\infty||u(t)||^p dt)^{(1/p)}<\infty$ for $p\in[1,\infty)$ and $||u||_{\Ls_\infty}=\sup_{t\geq0}||u(t)||<\infty$. When the subscript is omitted, it means any $p$ is allowed. Whenever the dimension of signals bears clarification, a superscript $m$ is added as $\Ls^m$. The extended space of $\Ls$, denoted $\Ls_e$, is defined as $\Set{u\mid u_\tau\in\Ls,\;\forall\tau\in[0,\infty)}$, where $u_\tau$ is the truncation of $u$ defined as $u_\tau(t) = u(t)$ for $0\leq t\leq\tau$ and $0$ otherwise. The set of real-valued functions with $r$ times continuously differentiable partial derivatives over their domain is denoted by $\C^r$. The $i^{\textrm{th}}$ element of a vector $x$ is denoted by $x^{(i)}$. The preimage of a function $f$ with respect to a subset $\Omega$ of its codomain is defined by $f^{-1}(\Omega){=}\{x{\mid} f(x) {\in} \Omega \}$. The transpose of $x{\in}\R^n$ is denoted by $x^\intercal$. The vector of ones in $\R^n$ is denoted by $1_n$ and $e_i$ contains only zeros, except for a $1$ in its $i^{th}$ component.

The interior, boundary, and closure of $\Omega\subset\R^n$ are denoted by $\Omega\degree$, $\del\Omega$, and $\bar{\Omega}$, respectively. The set of all compact subsets $\Omega\subset\R^n$ satisfying i) $\Omega\degree$ is connected and contains the origin, and ii) $\Omega=\overline{\Omega\degree}$, is denoted by $\mathfrak{R}^n$. 

This paper concerns $\Ls$ stability, which is defined next.

\begin{definition}[$\Ls$ stability{\citep{zames1966input}}]\label{def:finiteGainStability}
A mapping $\mathcal{G}:\Ls_e^m\mapsto\Ls_e^q$ is $\Ls$ finite-gain stable if there exist $\gamma_1,\gamma_2\geq0$ such that 
\begin{equation} \label{eq:finiteGainStability}
    ||(\mathcal{G}u)_\tau||_\Ls \leq \gamma_1||u_\tau||_\Ls + \gamma_2
\end{equation}
for all $u\in\Ls_e^m$ and $\tau\in[0,\infty)$. When a $\gamma_1\geq0$ is found, the $\Ls$ gain of $\mathcal{G}$ is less than or equal to $\gamma_1$. \qed
\end{definition}

\noindent Since $\mathcal{G}$ is a state-space model of a dynamic system in this paper, causality, $(\mathcal{G}u)_\tau=(\mathcal{G}u_\tau)_\tau$ holds. When inputs are constrained, Definition\;\ref{def:finiteGainStability} is modified as follows to only allow a subset of the input space.

\begin{definition}[Small-signal $\Ls$ stability{\citep[Def\;5.2]{khalil}}]\label{def:smallSignalFGstab}
The mapping $\mathcal{G}:\Ls_e^m\mapsto\Ls_e^q$ is $\Ls$ small-signal finite-gain stable if there exists $r_u>0$ such that \eqref{eq:finiteGainStability} is satisfied for all $u \in \Ls_e^m$ with $\sup_{0\leq t\leq \tau}||u(t)||\leq r_u$. \qed
\end{definition}

We use \ac{CPA} functions to search for storage and barrier functions. They are defined on  triangulation, described next. 

\begin{definition}[Affine independence{\citep{gieslRevCPA2013}}] \label{def:affDepVecs}
A set of vectors $\{x_0,\ldots,x_n\}$ in $\R^n$ is called affinely independent if $x_1-x_0,\ldots,x_n-x_0$ are linearly independent. \qed
\end{definition}

\begin{definition}[$n$-simplex {\citep{gieslRevCPA2013}}] \label{def:simplex}
An $n$-simplex is the convex combination of $n+1$ affinely independent vectors in $\R^n$, denoted $\sigma{=}\textrm{co}(\{x_j\}_{j=0}^n)$, where $x_j$'s are called vertices. \qed
\end{definition}

\noindent In this paper, simplex always refers to $n$-simplex. By abuse of notation, $\T$ will refer to both a collection of simplexes and the set of points in all the simplexes of the collection. 

\begin{definition} [Triangulation {\citep{gieslRevCPA2013}}] \label{def:triangulation}
A set $\T\in\mathfrak{R}^n$ is called a triangulation if it is a finite collection of $\mt$ simplexes, denoted $\T=\{\sigma_i\}_{i=1}^{\mt}$, and the intersection of any two simplexes in $\T$ is either a face or the empty set. 

The following two conventions are used throughout this paper for triangulations and their simplexes. Let $\T=\{\sigma_i\}_{i=1}^n$. Further, let $\{x_{i,j}\}_{j=0}^n$ be $\sigma_i$'s vertices, making $\sigma_i=\textrm{co}(\{x_{i,j}\}_{j=0}^n)$. The choice of $x_{i,0}$ in $\sigma_i$ is arbitrary unless $0\in\sigma_i$, in which case $x_{i,0}=0$. The vertices of the triangulation $\T$ that are in $\Omega\subseteq\T$ is denoted by $\mathbb{E}_\Omega$. \qed
\end{definition}

\begin{lemma}[{\!\!\citep[Rem.\;9]{gieslRevCPA2013}}]\label{lem:nablaLinear}
Consider the triangulation $\T=\{\sigma_i\}_{i=1}^{\mt}$, where $\sigma_i=\textrm{co}(\{x_{i,j}\}_{j=0}^n)$, and a set $\mathbf{W}=\left\{ W_x \right\}_{ x\in \Et } \subset \R$. Let 
$X_i\in\R^{n\times n}$ be a matrix that has $x_{i,j}-x_{i,0}$ as its $j$-th row, and $\bar{W}_i{\in}\R^n$ be a vector that has $W_{x_{i,j}}{-}W_{x_{i,0}}$ as its $j$-th element. The function $W(x)=(x-x_{i,0})^\intercal X^{-1}_i \bar{W}_i+W_{x_{i,0}}$ is the unique, CPA interpolation of $\textbf{W}$ on $\T$, satisfying $W(x)=W_x$, $\forall x\in\Et$. \qed
\end{lemma}

The Dini derivative of a CPA function $W$ at $x\in\R^n$ on the trajectories of $\dot{x}=g(x)$ is defined in terms of the limit superior over positive $h\in\R$,
\begin{align*}
    D^+W(x) = \textrm{lim\,sup}_{h\rightarrow0^+}\sfrac{(W(x+hg(x))-W(x))}{h},
\end{align*} which equals $\dot{W}$ on $\Omega\subset\R^n$ if $W\in\C^1(\Omega)$ \citep{gieslRevCPA2013}. Also, a continuous function $g(x)\in\R^n$ is piecewise $\C^2$ on a triangulation $\T=\{\sigma_i\}_{i=1}^{\mt}$, denoted $g\in\C^2(\T)$, if it is in $\C^2$ on $\sigma_i$ for all $i\in\IntSet_1^{\mt}$ \citep[Def.\;5]{me}. The following lemma, which will be used frequently, overbounds a $\C^2$ vector function on a simplex.
\begin{lemma}[{\!\!\citep[Thm.\;1]{gieslRevCPA2013},\citep[Prop.\;2.2,\;Lem.\;2.3]{giesl2012}}]\label{lem:upBound}
    Consider $\hat{\Omega}\in\mathfrak{R}^n$ and let $g:\hat{\Omega}\mapsto\R^n$ satisfy $g\in\C^2(\T)$ for some triangulation, $\T=\{\sigma_i\}_{i=1}^{\mt}$ of $\hat{\Omega}$. Then, for any $x\in\sigma_i=\textrm{co}(\{x_{i,j}\}_{j=0}^n)\in\T$,
    \begin{equation}
        ||g(x) - \sum_{j=0}^n\lambda_jg(x_{i,j})||_\infty \leq \underline{\beta}_i \sum_{j=0}^n \lambda_jc_{i,j},
    \end{equation}
where $\{\lambda_j\}_{j=0}^n\in\R$ is the set of unique coefficients satisfying $x=\sum_{j=0}^n\lambda_jx_{i,j}$ with $\sum_{j=0}^n\lambda_j=1$ and $0\leq\{\lambda_j\}_{j=0}^n\leq1$, and 
\begin{align*}
    c_{i,j}{=}& \frac{n}{2} ||x_{i,j} {-} x_{i,0}|| (\max_{k\in\IntSet_1^n} ||x_{i,k}{-}x_{i,0}|| {+} ||x_{i,j}{-}x_{i,0}||), \textrm{ and} \\
    \underline{\beta}_i \geq& \max_{p,q,r\in\IntSet_1^n} \max_{\xi\in\sigma_i} \left| \left. \sfrac{\partial^2 g(x)^{(p)}}{\partial x^{(q)}\partial x^{(r)}} \right|_{x=\xi} \right|. 
\end{align*}
\qed
\end{lemma}

One of the key contributions of this work is performing gain analysis for constrained nonlinear systems. Barrier functions will be a key tool allowing us to identify under which inputs the states will remain feasible. The following definition modifies the zeroing barrier function proposed in \citep{Ames2016} by  allowing $W$ to have positive time derivatives inside $\A_1$, which means the decrease condition is not required everywhere in $\A$.
\begin{definition}[Modified Barrier Function {\citep{Lavaei2024Lyap}}] \label{def:myBarrier} Consider the system \\$\dot{x}=g(x)$, $x\in\X\in\mathfrak{R}^n$, where $g:\X\mapsto\R^n$ is a Lipschitz map.
Let $\hat{\Omega},\A_1\in\mathfrak{R}^n$ satisfy $\A_1\subset\hat{\Omega}$. Further, let $W:\hat{\Omega}\mapsto\R$ be a Lipschitz function satisfying 
\begin{subequations}\label{eq:barrierDef}
\begin{alignat}{2}
    W(x) &> 0, \quad && \forall x\in\hat{\Omega}, \label{eq:WposDef} \\
    D^+W(x) &\leq -b_2, \quad && \forall x\in (\hat{\Omega}\backslash\A_1)\degree, \label{eq:dWdef}
\end{alignat}
\end{subequations}
\noindent with $b_2>0$. Let $\A$ be a sublevel set of $W$ for which $\A_1\subset\A\subseteq\hat{\Omega}$ holds. Then, the restriction of $W$ to $\A\degree$, that is $W:\A\degree\mapsto\R$, is a barrier function. \qed
\end{definition}
\begin{definition}[Robust Barrier Function] \label{def:robustBarrier}
Consider the system \\$\dot{x}=g(x,u)$, $x\in\X\in\mathfrak{R}^n$, $u\in\mathcal{U}\in\mathfrak{R}^m$ where $g:\X\mapsto\R^n$ is a Lipschitz map in $x$ for all $u\in\mathcal{U}$. Let $\hat{\Omega},\A_1\in\mathfrak{R}^n$ satisfy $\A_1\subset\hat{\Omega}$. The restriction of a function $W:\hat{\Omega}\mapsto\R$ to its sublevel set, $\A$, that is $W:\A\degree\mapsto\R$, for which $\A_1\subset\A\subseteq\hat{\Omega}$ holds, is a robust barrier function to $u\in\mathcal{U}$ if for each  $u\in\mathcal{U}$, it is a modified barrier function for the system $\dot{x}=g(x,u)$.
\end{definition}
\section{Problem Statement}\label{sc:problem}
Consider the constrained mapping $\mathcal{G}:\Ls^m_e\mapsto\Ls^q_e$ defined by $y=\mathcal{G}u$ and
\begin{flalign} \label{eq:ContAffineSystem}
    &\dot{x} = f(x)+G(x)u, \;x\in\X\in\mathfrak{R}^n, \;u\in\mathcal{U}\in\mathfrak{R}^m, \\
    &y = h(x),\ \forall t\in[0,\infty)\nonumber
\end{flalign}
where $f(0)=0$ and $h(0)=0$. Find the $\Ls_2$-gain on a bounded region around the origin and identify a bound on the input, $||u||$, ensuring that trajectories originating in that region remain in it.

This paper seeks a solution to the above problem. That is, we find an upper bound on the $\Ls_2$-gain so that together with a bound on the inputs, a subset of admissible states is rendered robust positive-invariant. This can be done by seeking storage and robust barrier functions that satisfy \ac{HJ} inequalities, but this would require verifying inequalities at an infinite number of points. By seeking \ac{CPA} functions, the inequalities need only be verified at finitely many points. However, this still entails optimization over nonlinear inequalities, which become numerous for refined triangulations. To pose more computationally tractable problems, we formulate convex relaxations of the \ac{HJ} inequalities.


\section{Main Results} \label{sc:mainResults} 

Gain is typically established using \ac{HJ} inequalities, like those found in \citep[Thm\;2]{van19922}, but these theorems typically apply over $\R^n$. Small-signal finite-gain $\Ls_2$ stability can be ensured similarly if the state remains in a neighborhood of the origin for all $t\geq 0$. 
We formulate the search for a \ac{CPA} function establishing such an \ac{HJ} inequality to ensure small-signal finite-gain $\Ls_2$ stability of constrained nonlinear systems. Exploiting the structure of \ac{CPA} functions, we put Lyapunov-like properties on a shell inside a triangulation to find a robust barrier function and the corresponding positive-invariant set. This works much like a control barrier function that ensures states remain in a feasible region, but is robust to bounded input disturbances and does not impose convergence to an equilibrium. Moreover, we characterize the set of inputs for which the small-gain properties hold. The theorems in this section formulate this search. The first one seeks a \ac{CPA} storage function $V$ to an \ac{HJ} inequality related to $\mathcal{L}_2$-gain.

\begin{thm} \label{thm:hj}
Consider the constrained mapping $\mathcal{G}:\Ls^m_e\mapsto\Ls^q_e$ defined by $y=\mathcal{G}u$ and \eqref{eq:ContAffineSystem}, 
where $f(0)=0$, $h(0)=0$, and $\bar{g}(0)=0$ for $\bar{g}(x)=||G(x)G^\intercal(x)||_\infty$. Suppose that $f,G,h\in\C^2(\T)$ for a triangulation, $\T =\{\sigma_i\}_{i=1}^{\mt}$, of a set $\Omega\in\mathfrak{R}^n$. There exist $\mathbf{V}=\{V_x\}_{x\in\Et}\subset\R$,  $\mathbf{L}=\{l_i\}_{i=1}^{\mt}\subset\R^n$, and $b_1,\gamma\in\R$ satisfying
\begin{subequations} \label{eq:hj} 
    \begin{alignat}{2}
        & \gamma > 0,  \; && \label{eq:hjGamma} \\
        & V_x \geq 0, \;&& \forall x\in\Et, \label{eq:hjVconstraint} \\
        & |{\nabla V}_i| \leq l_i, &&\forall i\in\IntSet_1^{\mt}, \label{eq:hjNablaConstraint} \\
        & H_{i,j} \leq -b_1, \;\; &&\forall i\in\IntSet_1^{\mt}, \; \forall j\in\IntSet_0^n, \; x\neq 0 \label{eq:hjhj}
        \end{alignat}
\end{subequations}
\noindent where $V_i$ is the affine interpolation of $\mathbf{V}$ on $\sigma_i\in\T$, $\nabla V_i$ is its gradient, and
\begin{align}
    \!\!\! H_{i,j} =& f(x_{i,j})^\intercal {\nabla V}_i + 0.5|| h(x_{i,j})||_2^2
    + (1_n^\intercal l_i\beta_i+0.5\tilde{\beta_i})c_{i,j}\nonumber\\
    &+ \frac{1}{2\gamma}(\bar{g}(x_{i,j}) + \bar{\beta}_i c_{i,j})(1_n^\intercal l_i)^2 \label{eq:Hij}\\
    c_{i,j}{=}& \frac{n}{2} ||x_{i,j} {-} x_{i,0}|| (\max_{k\in\IntSet_1^n} ||x_{i,k}{-}x_{i,0}|| {+} ||x_{i,j}{-}x_{i,0}||), \label{eq:betaAndc}\\
    \beta_i \geq& \max_{p,q,r\in\IntSet_1^n} \max_{\xi\in\sigma_i} \left| \left. \sfrac{\partial^2 f(x)^{(p)}}{\partial x^{(q)}\partial x^{(r)}} \right|_{x=\xi} \right|, \label{eq:beta} \\
    \tilde{\beta}_i \geq& \max_{q,r\in\IntSet_1^n} \max_{\xi\in\sigma_i} \left| \left. \sfrac{\partial^2 h^\intercal(x)h(x)}{\partial x^{(q)}\partial x^{(r)}} \right|_{x=\xi} \right|,\text{ and} \label{eq:betaTilde} \\
    \bar{\beta}_i \geq& \max_{q,r\in\IntSet_1^n} \max_{\xi\in\sigma_i} \left| \left. \sfrac{\partial^2 \bar{g}(x)}{\partial x^{(q)}\partial x^{(r)}} \right|_{x=\xi} \right|. \label{eq:betaBar}
\end{align}

\noindent If further $b_1 > 0$, then the \ac{HJ} inequality,
\begin{align}\label{eq:HJineq}
    \!\!\!\!\hat{H} \triangleq \nabla{V}^\intercal(x) f(x) + \frac{1}{2\gamma}||G^\intercal(x)\nabla{V}(x)||_2^2 + \frac{1}{2} ||h(x)||_2^2 \leq 0,
\end{align}
is satisfied for all $x\in\Omega\degree$.
\qed
\end{thm}

\begin{pf}
To see that \eqref{eq:hj} is feasible, observe that 
for any $\gamma$ and $V_x$ satisfying \eqref{eq:hjGamma}-\eqref{eq:hjVconstraint}, Lemma\;\ref{lem:nablaLinear} can be used to compute $l_i=|\nabla{V}_i|$, verifying \eqref{eq:hjNablaConstraint}. Each $c_{i,j}$ is finite because $\Omega\in\mathfrak{R}^n$ implies $\T$ is bounded. Finite $H_{i,j},\, \beta_i,\, \tilde{\beta}_i$,\, $\bar{\beta}_i$, and consequently $b_1$, can be found because $f,G,h\in\C^2(\T)$. So finite $b_1$ exists satisfying \eqref{eq:hjhj}, though it may be negative.

To show that \eqref{eq:hj} with $b_1> 0$ implies \eqref{eq:HJineq} on $\Omega\degree$, we begin bounding each term in \eqref{eq:HJineq} using Lemma\;\ref{lem:upBound}. This is possible because for any $x\in\Omega\degree$,
there is a simplex, $\sigma_i=\textrm{co}(\{x_{i,j}\}_{j=0}^n)$, such that $x=\sum_{j=0}^n\lambda_jx_{i,j}$, $0\leq\{\lambda_i\}_{i=0}^n\leq1$, and $\sum_{j=0}^n\lambda_j=1$.

The first term's bound in \eqref{eq:HJineq} was established in \citep[Thm.\;1]{gieslRevCPA2013}, and is reproduced here for completeness. Combining the fact that $\nabla{V}^\intercal f(x)=\nabla{V}_i^\intercal f(x)$ on $\sigma_i$, H\"{o}lder's inequality, \eqref{eq:hjNablaConstraint},\eqref{eq:beta}, and Lemma\;\ref{lem:upBound} yields
\begin{align}
    \nabla&{V}^\intercal f(x) = \nabla{V}_i^\intercal\left( \sum_{j=0}^n\lambda_jf(x_{i,j}) + f(x)-\sum_{j=0}^n\lambda_jf(x_{i,j})\right)\nonumber\\
    \leq&\sum_{j=0}^n\lambda_j \nabla{V}_i^\intercal f(x_{i,j}) + ||\nabla{V}_i||_1 ||f(x)-\sum_{j=0}^n\lambda_jf(x_{i,j})||_\infty\nonumber\\
    \leq&\sum_{j=0}^n\lambda_j \nabla{V}_i^\intercal f(x_{i,j}) + \beta_i 1_n^\intercal l_i \sum_{j=0}^n\lambda_jc_{i,j}. \label{eq:fbound}
\end{align}

The second term of \eqref{eq:HJineq} will now be re-formulated so it can be bounded in union with the rest.
For $x\in\sigma_i$, applying H\"{o}lder's inequality, submultiplicativity of norms, the fact that $||v||_\infty \leq ||v||_1$ for any $v\in\R^n$, and \eqref{eq:hjNablaConstraint} then show that
\begin{flalign}
    ||&G^\intercal(x)\nabla{V}||_2^2
    \leq ||\nabla{V}_i||_1 ||G(x)G^\intercal(x) \nabla{V}_i||_\infty \label{eq:vGGvBound} \\
    &\leq ||\nabla{V}_i||_1 \bar{g}(x) ||\nabla{V}_i||_\infty \leq ||\nabla{V}_i||_1^2 \bar{g}(x) \leq (1_n^\intercal l_i)^2 \bar{g}(x).  \nonumber
\end{flalign}

Note that $\bar{g}(x)$ and $\bar{h}(x) \triangleq h^\intercal(x)h(x)$ and the third term in \eqref{eq:HJineq} are non-negative scalar functions.  
Using Lemma\;\ref{lem:upBound} once with $g(x)=\bar{g}(x)$ and again with $g(x)=\bar{h}(x)$ together with the fact that $||v||_\infty=|v|\;\forall v\in\R$ yields
\begin{align}
    \frac{1}{2\gamma}\bar{g}(x) &\leq \sum_{j=0}^n \frac{1}{2\gamma}\left(\lambda_j\bar{g}(x_{i,j}) + \bar{\beta}_i\lambda_jc_{i,j}\right)\text{, and} \label{eq:gbarBound} \\
    \bar{h}(x) &\leq \sum_{j=0}^n \left( \lambda_j\bar{h}(x_{i,j}) + \tilde{\beta}_i\lambda_jc_{i,j}\right). \label{eq:hbarBound}
\end{align}

The upper bound on \eqref{eq:HJineq} can now be obtained. Summing \eqref{eq:fbound}, \eqref{eq:gbarBound}, and \eqref{eq:hbarBound} reveals that
$\hat{H}(x) \leq \sum_{j=0}^n \lambda_j H_{i,j}$, with $H_{i,j}$ given by \eqref{eq:Hij}. Finally, using \eqref{eq:hjhj} on $\sigma_i$ and the facts that 
$0\leq\{\lambda_i\}_{i=0}^n\leq1$ and $\sum_{j=0}^n\lambda_j=1$ yields 
\begin{equation}
    \hat{H}(x) \leq \sum_{j=0}^n \lambda_j H_{i,j} \leq -b_1 \sum_{j=0}^n \lambda_j = -b_1.
\end{equation}
Therefore, with $b_1>0$, \eqref{eq:HJineq} is verified at any $x\in\Omega\degree\setminus\{0\}.$ This also holds for $x=0$ because $f(0)=h(0)=\bar{g}(0)=0$ by assumption, $c_{i,0}=0$ by construction,  and $x_{i,0}=0$ if $0\in \sigma_i$. Hence, $H_{i,0}=0$, verifying \eqref{eq:HJineq} for all $x\in\Omega\degree$. \hfill $\blacksquare$
\end{pf}

In practice, even if $b_1>0$ is not found to satisfy Theorem\;\ref{thm:hj} on $\Omega$, there might be a sub-triangulation of $\Omega$ on which $H_{i,j}$ is negative at all its vertices. The sub-triangulated region constitutes a set where Theorem\;\ref{thm:hj} is satisfied.

Note that Theorem\;\ref{thm:hj} satisfies \eqref{eq:HJineq} only on a subset of $\R^n$ because otherwise, \eqref{eq:hj} would have an infinite number of constraints. We will use small-signal properties to make sure that the state does not escape that subset. To do so, we search for a \ac{CPA} robust barrier function separately by verifying the conditions of the following theorem.

\begin{thm} \label{thm:barrier}
Consider the system
\begin{flalign} \label{eq:barriersystem}
    &\dot{x} = f(x)+G(x)u, \;x\in\hat{\Omega}\in\mathfrak{R}^n, \\
    &u = \kappa(t),\ \forall t\in[0,\infty)\nonumber
\end{flalign}
where $f(0)=0$. Let $\T =\{\sigma_i\}_{i=1}^{\mt}$ be a triangulation of $\hat{\Omega}$. There exist $\mathbf{W}=\{W_x\}_{x\in\Et}\subset\R$, $\hat{\mathbf{L}}=\{\hat{l}_i\}_{i=1}^{\mt}\subset\R^n$, and $b_2,\hat{u}\in\R$ satisfying
\begin{subequations} \label{eq:barrier} 
    \begin{align}
        \hat{u} >& 0,\label{eq:barr_uhat} \\
        W_x >& 0,\ \forall x\in\Et, \label{eq:barr_Wconstraint} \\
        ||{\nabla W}_i||_1 \leq& \hat{l}_i,\ \forall i\in\IntSet_1^{\mt}, \label{eq:barr_NablaConstraint} \\
        D^+_{i,j}W \leq& -b_2,\ \forall i\in\I_1,\ \forall j\in\IntSet_0^n, \label{eq:barr_dw}
        \end{align}
\end{subequations}
where $D^+_{i,j}W = f(x_{i,j})^\intercal {\nabla W}_i + (1_n^\intercal \hat{l}_i)(\beta_i c_{i,j} + \hat{g}_i\hat{u})$, $\hat{g}_i=\max_{x\in\sigma_i}||G(x)||_\infty$,  $c_{i,j}$ is given in \eqref{eq:betaAndc}, and $\beta_i$ is given in \eqref{eq:beta}.  Suppose further that $b_2>0$ and there exists a sublevel set of $W(x)$, $\A\subseteq\hat{\Omega}$, and a second set, $\A_1\in\mathfrak{R}^n$, such that $\A_1\subset\A$ and $\partial\A_1\subseteq\{\partial\sigma_i\}_{i=1}^{\mt}$. 
Then, the restriction of $W(x)$ to $\A\degree$, $W:\A\degree\mapsto\R$, is a robust barrier function provided $||\kappa(t)||_\infty \leq \hat{u}$ for $t \geq 0$. \qed
\end{thm}

\begin{pf}
This proof parallels that of Theorem\;\ref{thm:hj}. The feasibility argument is nearly unchanged, so we focus on showing that \eqref{eq:barrier} with $b_2 > 0$ satisfies Definition\;\ref{def:myBarrier}.

Since $W$ is CPA, \eqref{eq:barr_Wconstraint} implies \eqref{eq:WposDef}. For any $x\in(\hat{\Omega}\backslash\A_1)\degree$, there is a simplex $\sigma_i=\textrm{co}(\{x_{i,j}\}_{j=0}^n)\subseteq\A_1$ 
satisfying $x=\sum_{j=0}^n\lambda_jx_{i,j}$ for $0\leq\{\lambda_i\}_{i=0}^n\leq1$ and $\sum_{j=0}^n\lambda_j=1$. On this simplex, \eqref{eq:barr_NablaConstraint} and \eqref{eq:barr_dw} imply that $\nabla{W}^\intercal{f(x)}\leq\sum_{j=0}^n\lambda_j(\nabla{W}^\intercal f(x)+1_n^\intercal \hat{l}_i \beta_i c_{i,j})$ following arguments similar to those that established a bound on $\nabla{V}^\intercal f(x)$ in Theorem\;\ref{thm:hj}'s proof.
Moreover, $\nabla{W}^\intercal G(x) u = \nabla{W}^\intercal_i G(x) u$ on $\sigma_i$. By H\"{o}lder's inequality $\nabla{W}^\intercal_i G(x) u\leq||\nabla{W}_i||_1||G(x)u||_\infty$. Submultiplicativity of norms, \eqref{eq:barr_NablaConstraint}, and the definitions of $\hat{g}_i$ and $\hat{u}$ imply $\nabla{W}^\intercal_i G(x) u\leq 1_n^\intercal\hat{l}_i\hat{g}_i\hat{u}$. Adding the two derived inequalities, we have that $D^+W(x)=\nabla{W}^\intercal\dot{x}\leq D^+_{i,j}W\leq -b_2$, which verifies \eqref{eq:dWdef}. \hfill $\blacksquare$
\end{pf}

Note that even if $b_2>0$ does not exist everywhere in $\mathcal{T}$, there might be a sub-triangulation containing $\A_1$ on which $D^+_{i,j} W$ is negative at all its vertices that are not in $\A_1\degree$. The restriction of $W$ to this sub-triangulation is a robust barrier function, but on the sub-triangulation, rather than $\hat{\Omega}$.

Using CPA functions $V$ and $W$, we can state the conditions for small-signal finite-gain $\Ls_2$ stability.

\begin{thm} \label{thm:total}
Suppose that Theorems\;\ref{thm:hj} and \ref{thm:barrier} both hold with $b_1,b_2>0$ and $\A\subseteq\Omega$. Then, \eqref{eq:ContAffineSystem} is small-signal finite-gain $\Ls_2$ stable in $\A\degree$ for all $u\in\Ls_e^m$ with
\begin{align*}
\sup_{0\leq t\leq\tau}||u(t)||_\infty \leq \hat{u}
\end{align*}
and the $\Ls_2$-gain is less than or equal to $\sqrt{\gamma}$. \qed
\end{thm}

\begin{pf}
Note that $\A$ is a sub-level set of W, which is a barrier function that is robust to inputs  $||u(t)||_\infty\leq\hat{u},\;\forall t\geq 0$. Positive-invariance of $\A\degree$ follows as a special case of \citep[Thm\;2.6]{giesl2015} because $b_2>0$ and $\A,\A_1\in\mathfrak{R}^n$ in Theorem\;\ref{thm:barrier}.

To bound the small-signal $\Ls_2-$gain, we must demonstrate that \eqref{eq:finiteGainStability} holds if $x(0)\in \A\degree$ and $||u(t)||_\infty\leq \hat{u}$. In fact, $||u(t)||_\infty\leq \hat{u}$ is only important because it assures that $x(t)\in \A\degree$, so analysis can be restricted to $\A\degree$. Considering \eqref{eq:ContAffineSystem}, $\dot{V}=\nabla{V}^\intercal (f(x) +G(x)u)$. By completing squares for the $\nabla{V}^\intercal G(x)u$ term using the two vectors $\sqrt{\gamma}u$ and $(1/\sqrt{\gamma})G(x)^\intercal\nabla{V}$ for $\gamma > 0$, we have
\begin{align} \label{eq:vdot_squares}
    \dot{V}= &\nabla{V}^\intercal f(x) + \frac{1}{2}\gamma ||u||_2^2 + \frac{1}{2\gamma} ||G(x)^\intercal\nabla{V}||_2^2 \nonumber \\ 
    & - \frac{\gamma}{2}||u -\frac{1}{\gamma}\nabla{V}^\intercal G(x)||_2^2.
\end{align}
Since $x(t)\in \A\degree\subseteq\Omega$, \eqref{eq:HJineq} can be substituted into \eqref{eq:vdot_squares} as 
\begin{equation*}
    \dot{V} \leq - \frac{1}{2}||y||_2^2 + \frac{1}{2}\gamma||u||_2^2 - \frac{\gamma}{2}||u -\frac{1}{\gamma}\nabla{V}^\intercal G(x)||_2^2.
\end{equation*}
This means $||y||_2^2 \leq \gamma ||u||_2^2 - 2\dot{V}$. Integrating both sides and considering that $V(x)\geq 0$ on $\Omega$, we obtain
\begin{equation*}
    \int_0^\tau ||y||_2^2 dt \leq \gamma \int_0^\tau ||u||_2^2 dt + 2V(x(0)).
\end{equation*}
Finally, taking square roots and applying the triangle inequality yields
\begin{equation*}
    ||y_\tau||_{\Ls_2} \leq \sqrt{\gamma} ||u_\tau||_{\Ls_2} + \sqrt{2V(x(0))},
\end{equation*}
which verifies \eqref{eq:finiteGainStability} with $\gamma_1=\gamma$ and $\gamma_2=\sqrt{2V(x(0))}$. \hfill $\blacksquare$
\end{pf}

Note that the three important characteristics of the system, namely $\gamma$, $\hat{u}$, and $\A$, found by Theorem\;\ref{thm:total} all depend on the choices for the triangulations and Hessian term's upper bounds used in Theorems\;\ref{thm:hj} and \ref{thm:barrier}. For instance, $\gamma$ is only an upper bound on the system's gain. Moreover, even if $\Omega$ in Theorem\;\ref{thm:hj} and $\hat{\Omega}$ are the same, their corresponding triangulations do not have to be the same since \eqref{eq:hj} and \eqref{eq:barrier} are solved independently.

\section{Efficient Algorithms} \label{sc:ContDesign}
While Theorem\;\ref{thm:total} presents sufficient conditions for finite-gain stability, it imposes non-convex constraints. This section proposes conservative, but convex, relaxations of the theorems, replacing them with  iterative \acp{SDP}. The following theorems formulate each iteration.

\subsection{Iterative \texorpdfstring{\acp{SDP} for Theorem\;\ref{thm:hj}}{SDPs for Theorem 2}}

While it is easy to obtain a feasible point of \eqref{eq:hj}, only feasible points with $b_1>0$ are of use and the tightest possible gain bound is desirable. Minimizing $-b_1$ subject to \eqref{eq:hj} constitutes a search for an appropriate point, but it is a non-convex problem. Once $b_1>0$, minimizing $\gamma$ subject to \eqref{eq:hj} constitutes a search for the tightest possible gain bound, but it is a non-convex problem too. Here, we establish more conservative, but convex, criteria to impose \eqref{eq:hj}, facilitating the required searches. Like most non-convex problems, it is unclear a priori where to begin searching, so two initialization heuristics are proposed.

\begin{initialization} \label{init:1}
    Assign $\gamma > 0$ and $V_x\geq 0$ for all $x\in\Et$ to satisfy \eqref{eq:hjGamma} and \eqref{eq:hjVconstraint}. To verify \eqref{eq:hjNablaConstraint}, compute $l_i=|\nabla{V}_i|$ for all $i\in\IntSet_{1}^{\mt}$ using Lemma\;\ref{lem:nablaLinear}. Use these to find $b_1\in\R$ by minimizing $-b_1$ subject to \eqref{eq:hjhj} on all simplexes. \qed
\end{initialization}

\begin{initialization} \label{init:2}
    Linearize \eqref{eq:ContAffineSystem} around the origin. Find $\gamma$ and the storage function, $V$, using the Bounded Real Lemma\;\citep[Lemma 3.17]{bao2007process}. Sample that function on the vertices to find $V_x$ satisfying \eqref{eq:hjVconstraint}. Compute $l_i=|\nabla{V}_i|$ for all $i\in\IntSet_{1}^{\mt}$ using Lemma\;\ref{lem:nablaLinear}. Assign $\gamma > 0$ to satisfy \eqref{eq:hjGamma}. Use these to find $b_1\in\R$ by minimizing $-b_1$ subject to \eqref{eq:hjhj} on all simplexes. \qed
\end{initialization}

Starting with \textit{any} feasible point of \eqref{eq:hj}, the following theorem establishes that a convex cost function can be iteratively minimized subject to an overbound of \eqref{eq:hj} through a series of \acp{SDP}. Towards verifying Theorem\;\ref{thm:hj}, $b_1>0$ can be sought by setting the cost function to $J(\mathbf{y})=-b_1$. Once a feasible point has been found with $b_1>0$, the constraints of Theorem\;\ref{thm:hjSDP} can be augmented with $b_1>0$ and a new objective can be chosen. In particular, selecting $J(\mathbf{y})=\gamma$ seeks the tightest possible gain bound.

\begin{thm} \label{thm:hjSDP}
Suppose that $\mathbf{y}=[\mathbf{V}, \mathbf{L}, b_1, \gamma]$ is a feasible point for \eqref{eq:hj}. Consider the following optimization.
\begin{subequations} \label{eq:SDPhj} 
    \begin{alignat}{2}
        &\mathbf{y}^\ast = \argmin_{\delta\mathbf{y}=[\delta\mathbf{V},\, \delta\mathbf{L}, \delta b_1, \, \delta\gamma]} \;\; && J(\mathbf{y}+\delta\mathbf{y})  \nonumber \\
        & \textrm{s.t.}  && \nonumber \\ 
        & \gamma + \delta \gamma > 0,\; && \label{eq:SDPhjGammaConstraint} \\
        & V_x + \delta V_x \geq 0, \;&& \forall x\in\Et, \label{eq:SDPhjVconstraint} \\
        & |\nabla{V}_i + \delta\nabla{V}_i| \leq l_i + \delta l_i, \;\;&&\forall i\in\IntSet_1^{\mt}, \label{eq:SDPhjNablaConstraint} \\
        & P_{i,j} \preceq 0, \;\; &&\forall i\in\IntSet_1^{\mt},\; j\in\IntSet_0^n, \; x \neq 0 \label{eq:SDPhjhj}
        \end{alignat}
\end{subequations}

\noindent where $\delta\nabla{V}_i$ in \eqref{eq:SDPhjNablaConstraint} equals $X_i^{-1}\delta \bar{V}_i$ as in Lemma\;\ref{lem:nablaLinear}, and 

\begin{equation}
    P_{i,j} = 
    \begin{bmatrix}
    \phi_{i,j} + b_1 + \delta b_1 && \ast \\
    \sqrt{e_{i,j}}1_n^\intercal(l_i+\delta l_i)  && -(\gamma + \delta \gamma)
    \end{bmatrix},
\end{equation}

\noindent with $\phi_{i,j} = f(x_{i,j})^\intercal(\nabla{V}_i + \delta\nabla{V}_i ) + 1_n^\intercal(l_i+\delta l_i)\beta_ic_{i,j} + \frac{1}{2}h^\intercal(x_{i,j})h(x_{i,j})+\frac{1}{2}\tilde{\beta}_ic_{i,j}$, and $e_{i,j}=\norm{G(x_{i,j})G^\intercal(x_{i,j})}_\infty/2$.
Then $\mathbf{y} + \delta\mathbf{y}$ is a feasible point for \eqref{eq:hj}, and $J(\mathbf{y} + \delta\mathbf{y})\leq J(\mathbf{y})$. \qed
\end{thm}
\begin{pf}
To see \eqref{eq:SDPhj}'s feasibility, observe that $\delta\mathbf{y}{=}0$ satisfies \eqref{eq:SDPhj} since in this case, \eqref{eq:SDPhj} is equivalent to \eqref{eq:hj} . Substitution reveals that \eqref{eq:SDPhjGammaConstraint}--\eqref{eq:SDPhjhj} imply \eqref{eq:hjGamma}--\eqref{eq:hjhj} for $\mathbf{y}{+}\delta\mathbf{y}$. Note that \eqref{eq:SDPhjhj} is implied by Schur Complement\citep[Ch\;2]{lmiBook}. Finally, $J(\mathbf{y} {+} \delta \mathbf{y}) {\leq} J(\mathbf{y})$ because otherwise $\delta\mathbf{y}{=}0$ would be a better, feasible solution. \hfill $\blacksquare$
\end{pf}

\subsection{Iterative \texorpdfstring{\acp{SDP} for Theorem\;\ref{thm:barrier}}{SDPs for Theorem 3}}
Having a feasible point of \eqref{eq:barrier}, a cost function can be minimized while imposing a more conservative version of \eqref{eq:barrier} iteratively through a series of \acp{SDP}. This approach requires a feasible point of \eqref{eq:barrier}. Two methods for obtaining such a point are given next.

\begin{initialization} \label{init:3}
    Assign $\hat{u} > 0$ and $W_x\leq 0$ for all $x\in\Et$ to satisfy \eqref{eq:barr_uhat} and \eqref{eq:barr_Wconstraint}. Compute $\hat{l}_i=|\nabla{W}_i|$ or all $i\in\IntSet_{1}^{\mt}$ using Lemma\;\ref{lem:nablaLinear}. This verifies \eqref{eq:barr_NablaConstraint}. Using the computed values, find $b_2\in\R$ by calculating $\min -b_2$ subject to \eqref{eq:barr_dw} on all simplexes.  \qed
\end{initialization}

\begin{initialization} \label{init:4}
    Linearize \eqref{eq:ContAffineSystem} around the origin. Design an LQR controller and find its corresponding quadratic Lyapunov function. Sample that function on the vertices to find $W_x$ satisfying \eqref{eq:barr_Wconstraint}. Compute $\hat{l_i}=|\nabla{W}_i|$ or all $i\in\IntSet_{1}^{\mt}$ using Lemma\;\ref{lem:nablaLinear}. Assign $\hat{u} > 0$ to satisfy \eqref{eq:barr_uhat}. Using the computed values, find $b_2\in\R$ by calculating $\min -b_2$ subject to \eqref{eq:barr_dw} on all simplexes. \qed
\end{initialization}

The following theorem formulates each step of the iterative improvement of the cost function in \eqref{eq:barrier} as an \ac{SDP}. 

\begin{thm} \label{thm:barrierSDP}
    Suppose that $\mathbf{y}=[\mathbf{W},\; \hat{\mathbf{L}},\; b_2, \; \hat{u}]$ is a feasible point for \eqref{eq:barrier}.
    Consider the following optimization
    \begin{subequations} \label{eq:SDPbarrier} 
    \begin{alignat}{2}
        \mathbf{y}^\ast &= \argmin_{\delta\mathbf{y}=[\delta\mathbf{W},\; \delta\hat{\mathbf{L}},\; \delta b_2, \; \delta\hat{u}]} \;\; && J(\mathbf{y} + \delta\mathbf{y})  \nonumber \\
        \textrm{s.t.}\;\;& \hat{u} + \delta\hat{u} > 0,  \; && \label{eq:SDPbarr_uhat} \\
        & W_x + \delta W_x > 0, \;&& \forall x\in\Et, \label{eq:SDPbarr_Wconstraint} \\
        & |{\nabla W}_i + \delta{\nabla W}_i  | \leq \hat{l}_i + \delta \hat{l}_i, \;&&\forall i\in\IntSet_1^{\mt}, \label{eq:SDPbarr_NablaConstraint} \\
        & Q_{i,j} \preceq 0, \;\; &&\forall i\in\I_1, \; \forall j\in\IntSet_0^n, \label{eq:SDPbarr_dw}
        \end{alignat}
\end{subequations}

where $\delta\nabla{V}_i$ in \eqref{eq:SDPbarr_NablaConstraint} equals $X_i^{-1}\delta \bar{V}_i$ as in Lemma\;\ref{lem:nablaLinear}, and 

\begin{equation}
    Q_{i,j} = 
    \begin{bmatrix}
    \varphi_{i,j} + b_2 + \delta b_2 && \ast && \ast \\
    1_n^\intercal\delta\hat{l}_i  && -2 && \ast \\
    \hat{g}_i\delta\hat{u} && 0 && -2
    \end{bmatrix},
\end{equation}

\noindent with $\varphi_{i,j} = f(x_{i,j})^\intercal(\nabla{W}_i + \delta\nabla{W}_i ) + 1_n^\intercal(l_i+\delta l_i)(\beta_ic_{i,j}+\hat{g}_i\hat{u}) + 1_n^\intercal\hat{l}_i\hat{g}_i\delta\hat{u}$. Then $\mathbf{y} + \delta\mathbf{y}$ is a feasible point for \eqref{eq:hj}, and $J(\mathbf{y} + \delta\mathbf{y})\leq J(\mathbf{y})$. \qed
\end{thm}

\begin{pf}
    The proof follows similarly to that of Theorem\;\ref{thm:hj}'s. To see that \eqref{eq:SDPbarr_dw} implies \eqref{eq:barr_dw}, note that $u^\intercal v {\leq} 0.5(u^\intercal u {+}v^\intercal v)$ for any vectors $u,v$ with the same dimension. Applying this fact and using Schur Complement, \eqref{eq:SDPbarr_dw} is obtained. \hfill $\blacksquare$
\end{pf}

\subsection{Algorithm 1: The Combined Search for Gain and Small-Signal Bounds}

Note that $b_1,b_2>0$ are needed in Theorems\;\ref{thm:hj} and \ref{thm:barrier}. However, the feasible initializations we provided for them don't guarantee that. Moreover, the smallest value for the system gain and largest $\hat{u}$ are sought. Algorithm\;\ref{alg:strategy} begins the initializations and seeks to verify the criteria of inequalities\;\ref{eq:SDPhj} and \ref{eq:SDPbarrier}. If Algorithm\;\ref{alg:strategy} terminates, it has found \ac{CPA} storage and robust barrier functions establishing a bound, $\gamma$, on the $\mathcal{L}_2$-gain of system \eqref{eq:ContAffineSystem} over a finite region, $\mathcal{A}$, and a bound on the magnitude of inputs, $\hat{u}$, that ensures trajectories starting in $\mathcal{A}$ remain in $\mathcal{A}$.

Algorithm\;\ref{alg:strategy} begins with Initialization\;\ref{init:1} or \ref{init:2} to find feasible points for inequalities\;\ref{eq:SDPhj} and \ref{eq:SDPbarrier}. Then, using $J=-b_1$ and $J=-b_2$ in inequalities\;\ref{eq:SDPhj} and \ref{eq:SDPbarrier}, respectively, iteratively increases their initial values until $b_1,b_2>0$ are found. At this point, a storage function and an upper bound on the system's gain are found on $\Omega$. The iterative search for smaller $\gamma$ is then pursued by setting $J=\gamma$ while keeping $b_1>0$ in Theorem\;\ref{thm:hjSDP}. Similarly, the largest $\hat{u}$ is sought by setting $J=-\hat{u}$ while keeping $b_2>0$ in Theorem\;\ref{thm:barrierSDP}.

Note that if $b_1$ ($b_2$) is negative at line\;\ref{line:repeat1} (line\;\ref{line:repeat2}), a positive $\hat{b}_1$ ($\hat{b}_2$) can be sought on a sub-triangulation. 


\begin{algorithm}[htbp]
	\caption{$\Ls_2$-Stability Analysis}
    \label{alg:strategy}
	\begin{algorithmic}[1]
	   \Require System \eqref{eq:ContAffineSystem}, the sets $\Omega,\hat{\Omega},\A_1\in\mathfrak{R}^n$ and their triangulations
	   \Ensure $\gamma$, $\hat{u}$, and $\A$
          \State Find a feasible $\mathbf{y}$ satisfying Theorem\;\ref{thm:hj} using Initialization\;\ref{init:1} or \ref{init:2}
          \State Let $J=-b_1$ in Theorem\;\ref{thm:hjSDP}
          \Repeat{}
            \State Solve \eqref{eq:SDPhj} \label{line:repeat1}
          \Until{ $b_1 > 0$ is found }
          \State Add the constraint $b_1 + \delta b_1>0$ to \eqref{eq:SDPhj} \Comment{To keep $b_1$ positive} \label{line:b1found}
          \State Let $J=\gamma$ in Theorem\;\ref{thm:hjSDP} \Comment{To minimize $\gamma$}
          \Repeat{}
            \State Solve \eqref{eq:SDPbarrier} \label{line:repeat2}
          \Until{ $\gamma$ is not decreasing anymore } \Comment{At this point, a storage function and an upper bound on the system's gain, $\gamma$, are found on $\Omega$} \label{line:gammaFound}
	   \State Find a feasible $\mathbf{y}$ satisfying Theorem\;\ref{thm:barrier} using Initialization\;\ref{init:3} or \ref{init:4}
	   \Repeat{}
	        \State Solve \eqref{eq:SDPbarrier}
	   \Until{ $b_2>0$ is found} \label{line:b2found}
          \State Add the constraint $b_2 + \delta b_2 >0$ to \eqref{eq:SDPbarrier} \Comment{To keep $b_2$ positive}
          \State Let $J=-\hat{u}$ in Theorem\;\ref{thm:barrierSDP} \Comment{To maximize $\hat{u}$}
          \Repeat{}
            \State Solve \eqref{eq:SDPbarrier}
          \Until{ $\hat{u}$ is not increasing anymore }
          \State Find $\A$ satisfying Theorems\;\ref{thm:barrier} and \ref{thm:total} \label{line:findA}
          \State Return $\gamma$, $\hat{u}$, and $\A$
	\end{algorithmic}
\end{algorithm}

\subsection{Triangulation Refinement}
If Algorithm\;\ref{alg:strategy} stagnates before finding a positive value for $b_1$ or $b_2$, refining the triangulation may help. By refining, the error bounds used in Taylor's Theorem get tighter. Also, it adds more parameters to $V,W$, making it possible for them to capture more complex behaviors. If the refined triangulation includes all the vertices of the coarser one, the $\mathbf{y}$ found on the coarser triangulation can be used to initialize the \acp{SDP} on the finer one. However, it is important to strike a balance between triangulation refinement and computational complexity. The development of adaptive refinement schemes, which target the simplexes that most limit $b_1$ and $b_2$, would facilitate this. Likewise, adaptive refinement and approximation schemes would be needed for non-traingulable sets. The exploration of advanced triangulation design is left to future work.

\section{Numerical Simulation}
Consider $\dot{x}^{(1)} = x^{(2)}$, $\dot{x}^{(2)} = - \sin x^{(1)} - x^{(2)} + x^{(2)} u$ with $h(x) = x^{(2)}$. For this system, $||G(x)G^\intercal(x)||_\infty = (x^{(2)})^2$ and $||G(x)||_\infty \leq x^{(2)}$. Note that $G(0)=0$, as required by Theorem\;\ref{thm:hj}. The objective is to establish small-signal $\Ls_2$ stability.

\begin{figure}[ht]
    \centering
    \begin{subfigure}[t]{0.9\linewidth}
	\centering
	\includegraphics[width=6cm]{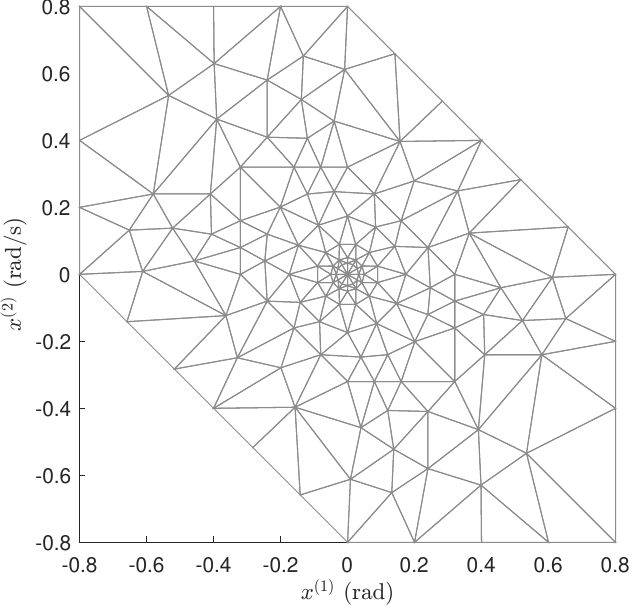}
	\caption{The set $\Omega$ and its triangulation on which the storage function and $\gamma$ were found} \label{fig:triangulation1}
    \end{subfigure}
    \\
    \begin{subfigure}[t]{0.9\linewidth}
	\centering
	\includegraphics[width=7cm]{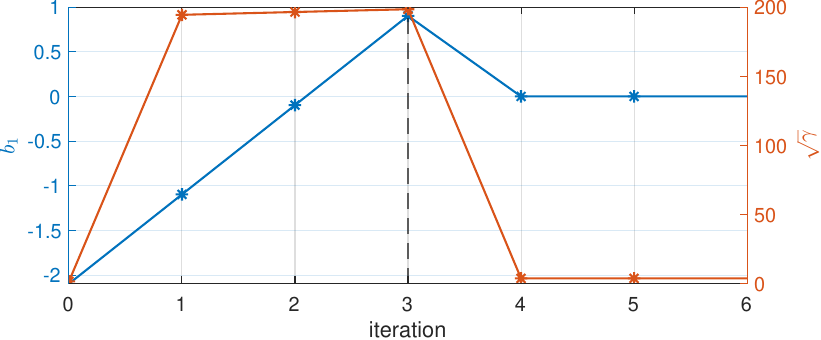}
	\caption{The $b_1$ and $\sqrt{\gamma}$ values produced by the iterative algorithm. After three iterations, a positive $b_1$ was found. Subsequent steps minimized $\gamma$ while keeping $b_1$ positive.} \label{fig:b1GammaProgress}
    \end{subfigure}
    \caption{Finding a storage function and the corresponding  gain, $\sqrt{\gamma}$}\label{fig:storage}
\end{figure}

The iterative procedure in Algorithm\;\ref{alg:strategy} is implemented for this example. The set $\Omega\in\mathfrak{R}^n$ and its triangulation on which a storage function and $\gamma$ were found are depicted in Fig\;\ref{fig:triangulation1}. It took three steps to reach $b_1=0.90$ from its initial $b_1=-2.10$ value found by initialization\;\ref{init:2}. At this point, $\gamma$ was $39403$. Then, by forcing $b_1$ to be positive, $\gamma$ was minimized iteratively. It reached $\gamma=14.80$ after only one iteration and stagnated, thus $\sqrt{\gamma}=3.85$ was obtained. At this point, $b_1$ was $0.0001$, which means a storage function and the corresponding gain, $\sqrt{\gamma}$ were found on $\Omega$. The obtained sequences of $b_2$ and $\sqrt{\gamma}$ are given in Fig.\;\ref{fig:b1GammaProgress}.

\begin{figure}[ht]
    \centering
    \begin{subfigure}[t]{0.9\linewidth}
	\centering
	\includegraphics[width=6cm]{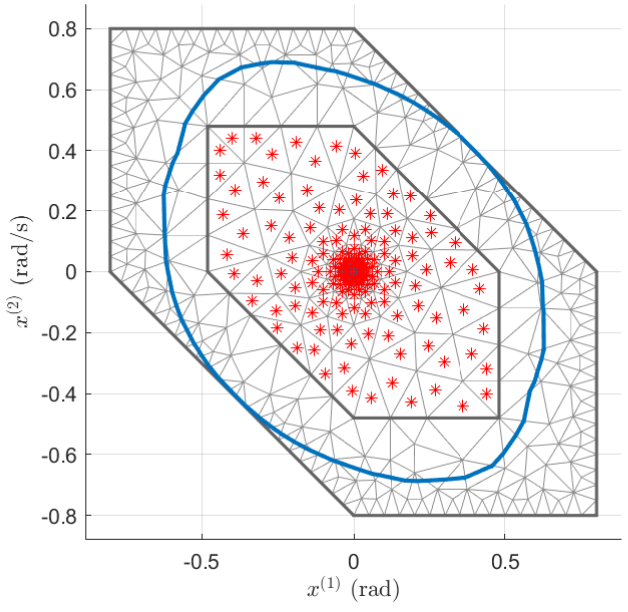}
	\caption{The sets $\hat{\Omega}$ and $\A_1$ satisfying $\A_1\subset\hat{\Omega}$ and the triangulation on which a robust barrier function and $\hat{u}$ were found. The boundary of a sub-level set $\A$ satisfying $\A_1\subset\A\subset\hat{\Omega}$ is in blue and the boundaries of $\hat{\Omega}$ and $\A_1$ are in dark gray. The simplexes with red asterisks are the ones on which $D^+W_{i,j}$ was not negative for at least one of their vertices.} \label{fig:triangulation2}
    \end{subfigure}
    \\
    \begin{subfigure}[t]{0.9\linewidth}
	\centering
	\includegraphics[width=7cm]{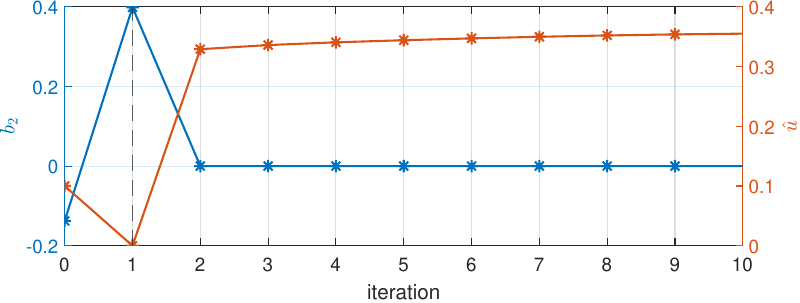}
	\caption{The $b_2$ and $\hat{u}$ values produced by the iterative algorithm. After one iteration, a positive $b_2$ was found. Subsequent steps maximized $\hat{u}$ while keeping $b_2$ positive.} \label{fig:b2uHatProgress}
    \end{subfigure}
    \caption{Finding a robust barrier function, the corresponding  $\hat{u}$, and the level-set $\A$}\label{fig:barrier}
\end{figure}

Next, a robust barrier function and $\hat{u}$ were sought to find a region from which the state would not escape. The boundaries of the sets $\hat{\Omega}$ and $\A_1$, where $\A_1\subset\hat{\Omega}$, are in dark gray in Fig.\ref{fig:triangulation2}. Note that $\hat{\Omega}$ in Fig.\ref{fig:triangulation2} equals $\hat{\Omega}$ but its triangulation is different. Although Theorem\;\ref{thm:barrier} does not have any requirement for triangulation of $\hat{\Omega}$ other than that it should cover $\partial\A_1$ by the faces of some simplexes, having finer simplexes close to $\partial\hat{\Omega}$ helped to find larger level sets. The simplexes marked by red asterisks in Fig.\;\ref{fig:triangulation2} denote the ones on which $D^+_{i,j}W$ was not negative on at least one of their vertices. It took only one step to get a positive $b_2$ after initializing using Initialization\;\ref{init:4}. At this point, $\hat{u}$ was $10^{-5}$. Then, by keeping $b_2$ positive, $\hat{u}$ was maximized iteratively until it stagnated at $\hat{u}=0.36$ after nine iterations. The $b_2$ and $\hat{u}$'s sequences are given in Fig.\;\ref{fig:b2uHatProgress}. The boundary of $\A$, the largest level set of the obtained robust barrier function in $\hat{\Omega}$, is depicted in blue in Fig.\;\ref{fig:triangulation2}. 

\clearpage
Finally, since $\Omega=\hat{\Omega}$ in this example, Theorem\;\ref{thm:total} implies that inside $\A$, small-signal finite-gain $\Ls_2$ stability holds with $\sqrt{\gamma}=3.85$ and $\hat{u}=0.36$.

\section{Conclusion}
This paper presented new criteria employing \ac{CPA} function approximations and error bounds in order to analyse the small-signal $\Ls_2$-gain of constrained, nonlinear state-space realizations. While finding the tightest possible bound is naturally a nonlinear, non-convex feasibility problem, iterative convex overbounding was used to verify the criteria through a sequence of well-posed \acp{SDP}. The resulting gain bounds and small signal requirements will be conservative due to the finite level of triangulation refinement and because iterative convex overbounding entails conservativeness. However, it provides practical algorithms to seek out the $\Ls_2$-gain of nonlinear systems, where few other approaches exist.

\section{Acknowledgements}
This work is supported by ONR under award \#N00014-23-1-2043.

\bibliography{BibliographyShort}

\begin{thebibliography}{10}
\expandafter\ifx\csname url\endcsname\relax
  \def\url#1{\texttt{#1}}\fi
\expandafter\ifx\csname urlprefix\endcsname\relax\def\urlprefix{URL }\fi
\expandafter\ifx\csname href\endcsname\relax
  \def\href#1#2{#2} \def\path#1{#1}\fi

\bibitem{boyd1987comparison}
S.~Boyd, J.~Doyle, Comparison of peak and {RMS} gains for discrete-time systems, Systems \& Control letters 9~(1) (1987) 1--6.

\bibitem{zames1981optimalsensitivity}
G.~Zames, Feedback and optimal sensitivity: Model meference transformations, multiplicative seminorms, and approximate inverses, IEEE Transactions on Automatic Control 26 (1981) 301--320.

\bibitem{zames1983minmax}
G.~Zames, B.~A. Francis, Feedback, minmax sensitivity, and optimal robustness, IEEE Transactions on Automatic Control 28 (1983) 585--601.

\bibitem{zames1966input}
G.~Zames, On the input-output stability of time-varying nonlinear feedback systems part one: Conditions derived using concepts of loop gain, conicity, and positivity, IEEE transactions on automatic control 11~(2) (1966) 228--238.

\bibitem{zhou1998essentials}
K.~Zhou, J.~C. Doyle, Essentials of robust control, Vol. 104, Prentice hall Upper Saddle River, NJ, 1998.

\bibitem{doyle1988state}
J.~Doyle, K.~Glover, P.~Khargonekar, B.~Francis, State-space solutions to standard {$\mathcal{H}_2$} and {$\mathcal{H}_\infty$} control problems, in: 1988 American Control Conference, IEEE, 1988, pp. 1691--1696.

\bibitem{boyd1989bisection}
S.~Boyd, V.~Balakrishnan, P.~Kabamba, A bisection method for computing the {$\mathcal{H}_\infty$} norm of a transfer matrix and related problems, Mathematics of Control, Signals and Systems 2~(3) (1989) 207--219.

\bibitem{bruinsma1990fast}
N.~Bruinsma, M.~Steinbuch, A fast algorithm to compute the {$\mathcal{H}_\infty$}-norm of a transfer function matrix, Systems \& Control Letters 14~(4) (1990) 287--293.

\bibitem{guo1988recursive}
L.~Guo, L.~Xia, Y.~Liu, Recursive algorithm for the computation of the {$\mathcal{H}_\infty$}-norm of polynomials, IEEE Transactions on Automatic Control 33~(12) (1988) 1154--1157.

\bibitem{scherer1990h}
C.~Scherer, {$\mathcal{H}_\infty$}-control by state-feedback and fast algorithms for the computation of optimal {$\mathcal{H}_\infty$}-norms, IEEE Transactions on Automatic Control 35~(10) (1990) 1090--1099.

\bibitem{lin1999computation}
W.-W. Lin, C.-S. Wang, Q.-F. Xu, On the computation of the optimal {$\mathcal{H}_\infty$} norms for two feedback control problems, Linear Algebra and its Applications 287~(1-3) (1999) 223--255.

\bibitem{gahinet1992numerical}
P.~Gahinet, P.~Apkarian, Numerical computation of the {$\mathcal{L}_\infty$} norm revisited, in: Proceedings of the 31st IEEE Conference on Decision and Control, IEEE, 1992, pp. 2257--2258.

\bibitem{gattami2015simple}
A.~Gattami, B.~Bamieh, Simple covariance approach to {$\mathcal{H}_\infty$} analysis, IEEE Transactions on Automatic Control 61~(3) (2015) 789--794.

\bibitem{benner2018}
P.~Benner, T.~Mitchell, Faster and more accurate computation of the {$\mathcal{H}_\infty$} norm via optimization, SIAM Journal on Scientific Computing 40~(5) (2018) A3609--A3635.
\newblock \href {https://doi.org/10.1137/17M1137966} {\path{doi:10.1137/17M1137966}}.

\bibitem{xia2020sector}
M.~Xia, P.~Gahinet, N.~Abroug, C.~Buhr, E.~Laroche, Sector bounds in stability analysis and control design, International Journal of Robust and Nonlinear Control 30~(18) (2020) 7857--7882.

\bibitem{van19922}
A.~J. Van Der~Schaft, {$\mathcal{L}_2$}-gain analysis of nonlinear systems and nonlinear state feedback {$\mathcal{H}_\infty$} control, IEEE Transactions on Automatic Control 37~(6) (1992) 770--784.

\bibitem{ball1996viscosity}
J.~A. Ball, J.~W. Helton, Viscosity solutions of hamilton-jacobi equations arising arising in nonlinear {$\mathcal{H}_\infty$} control, in: control, J. Math. Systems Estim. Control, AMS, 1996, pp. 1--42.

\bibitem{helton1999extending}
J.~W. helton, M.~R. James, Extending {$\mathcal{H}_\infty$} Control to Nonlinear Systems: Control of Nonlinear Systems to Achieve Performance Objectives, SIAM, 1999.

\bibitem{zhang2012performance}
H.~Zhang, P.~M. Dower, Performance bounds for nonlinear systems with a nonlinear {$\mathcal{L}_2$}-gain property, International Journal of Control 85~(9) (2012) 1293--1312.

\bibitem{summers2013quantitative}
E.~Summers, A.~Chakraborty, W.~Tan, U.~Topcu, P.~Seiler, G.~Balas, A.~Packard, Quantitative local {$\mathcal{L}_2$}-gain and reachability analysis for nonlinear systems, International Journal of Robust and Nonlinear Control 23~(10) (2013) 1115--1135.

\bibitem{giesl2012}
P.~Giesl, S.~Hafstein, Construction of {L}yapunov functions for nonlinear planar systems by linear programming, J. Math. Anal. Appl. 388~(1) (2012) 463--479.

\bibitem{gieslRevCPA2013}
P.~A. Giesl, S.~F. Hafstein, Revised {CPA} method to compute {L}yapunov functions for nonlinear systems, J Math Analysis \& Apps 410~(1) (2014) 292--306.

\bibitem{giesl2015}
P.~Giesl, S.~Hafstein, \href{https://doi.org/10.1137/140988802}{Computation and verification of {L}yapunov functions}, SIAM J on Appl Dyn Sys 14~(4) (2015) 1663--1698.
\newblock \href {http://arxiv.org/abs/https://doi.org/10.1137/140988802} {\path{arXiv:https://doi.org/10.1137/140988802}}, \href {https://doi.org/10.1137/140988802} {\path{doi:10.1137/140988802}}.
\newline\urlprefix\url{https://doi.org/10.1137/140988802}

\bibitem{khalil}
H.~Khalil, \href{https://books.google.com/books?id=t\_d1QgAACAAJ}{Nonlinear Systems}, Pearson Edu, Prentice Hall, 2002.
\newline\urlprefix\url{https://books.google.com/books?id=t\_d1QgAACAAJ}

\bibitem{me}
R.~Lavaei, L.~Bridgeman, Simultaneous controller and {L}yapunov function design for constrained nonlinear systems, in: 2022 American Control Conference (ACC), IEEE, 2022, pp. 4909--4914.

\bibitem{Ames2016}
A.~D. Ames, X.~Xu, J.~W. Grizzle, P.~Tabuada, Control barrier function based quadratic programs for safety critical systems, IEEE Trans Aut Ctrl 62~(8) (2016) 3861--3876.

\bibitem{Lavaei2024Lyap}
R.~Lavaei, L.~J. Bridgeman, Systematic, lyapunov-based, safe and stabilizing controller synthesis for constrained nonlinear systems, IEEE Transactions on Automatic Control (2023) 1--12\href {https://doi.org/10.1109/TAC.2023.3302789} {\path{doi:10.1109/TAC.2023.3302789}}.

\bibitem{bao2007process}
J.~Bao, P.~L. Lee, Process control: the passive systems approach, Springer Science \& Business Media, 2007.

\bibitem{lmiBook}
S.~Boyd, L.~{El~{G}haoui}, E.~Feron, V.~Balakrishnan, Linear Matrix Inequalities in System and Control Theory, Vol.~15 of Studies in Applied Mathematics, {SIAM}, Philadelphia, PA, 1994.

\end{thebibliography}

\begin{acronym}
\acro{SDP}{semi-definite program}
\acro{MPC}{model predictive control}
\acro{CLF}{control Lyapunov function}
\acro{CBF}{control barrier function}
\acro{CPA}{continuous piecewise affine}
\acro{QP}{quadratic programming}
\acro{DP}{dynamic programming}
\acro{ROA}{region of attraction}
\acrodefplural{ROA}{regions of attraction}
\acro{PWA}{piecewise affine}
\acro{LMI}{linear matrix inequality}
\acrodefplural{LMI}{linear matrix inequalities}
\acro{BMI}{bilinear matrix inequality}
\acrodefplural{BMI}{bilinear matrix inequalities}
\acro{EMPC}{explicit model predictive control}
\acro{HJ}{Hamilton-Jacobi}
\end{acronym}

\end{document}